\documentclass[10pt,letterpaper]{article}

\usepackage{opex3}

\usepackage[draft]{hyperref}

\usepackage{bm}
\usepackage{amsmath}
\usepackage{amssymb}

\usepackage{graphicx,color}
\usepackage{epsfig}

\newcommand{\ket}[1]{\ensuremath{|\,{#1}\,\rangle}}
\newcommand{\bra}[1]{\ensuremath{\langle\,{#1}\,|}}

\newcommand{\itgf}[1]{\ensuremath{\int\!\!d{#1}\,}}
\newcommand{\sinc}{\ensuremath{\mbox{\hspace{1.3pt}sinc}\,}}

\begin{document}

\title{Experimental quantum tomography of photonic qudits via mutually unbiased basis}

\author{G. Lima,$^{1,2,*}$ L. Neves,$^{1,2}$, R. Guzm\'{a}n,$^{1,3}$, E.~S.~G\'{o}mez,$^{1,2}$ W.~A.~T.~Nogueira,$^{1,2}$ A.~Delgado,$^{1,2}$ A. Vargas,$^{1,3}$  and C. Saavedra$^{1,2}$}

\address{
$^1$Center for Optics and Photonics, Universidad de Concepci\'{o}n, Casilla 4016, Concepci\'{o}n, Chile\\
$^2$Departamento de F\'{\i}sica, Universidad de Concepci\'{o}n, Casilla 160-C, Concepci\'{o}n, Chile \\
$^3$Departamento de Ciencias F\'{\i}sicas, Universidad de La Frontera, Temuco, Casilla 54-D, Chile \\
$^*$Corresponding author: glima@udec.cl}

\begin{abstract}
We present the experimental quantum tomography of 7- and 8-dimensional quantum systems based on projective measurements in the mutually unbiased basis (MUB-QT). One of the advantages of MUB-QT is that it requires projections from a minimal number of bases to be performed. In our scheme, the higher dimensional quantum systems are encoded using the propagation modes of single photons, and we take advantage of the capabilities of amplitude- and phase-modulation of programmable spatial light modulators to implement the MUB-QT.
\end{abstract}


\section{Introduction}

Quantum tomography (QT) is a process that allows the state reconstruction of physical systems. An arbitrary $D$-dimensional quantum system (qudit-$D$) is represented by a positive semidefinite, unit-trace Hermitian operator, which requires only $D^2-1$ independent real numbers for its specification. In the standard QT \cite{Fano,James}, when $D=2^N$ (i.e., a composite system of $N$ qubits), these parameters are determined by projecting the density operator onto completely factorized bases \cite{James,Thew}. The projective measurements used form a convenient set from the point of view of experimental implementation but, unfortunately, it does not allow the optimal quantum state reconstruction. Even when only local projective measurements are considered, there is another set that provides better performance for the QT \cite{Langford}. However, as it was shown by Wooters and Fields \cite{Wooters2}, the optimal way for reducing the effects of the inherent statistical errors in the QT can only be achieved with measurements on mutually unbiased bases (MUBs), where any two vectors of different bases have the same overlap's absolute value. This has been demonstrated experimentally in \cite{Steinberg} for polarization-entangled states and we refer to it as MUB-QT. In the MUB-QT, the projective measurements considered cannot be described in terms of factorized bases.

Besides, while reconstructing qudit states, it is necessary to work with an overcomplete set of measurements that allows more accurate normalization for converting the recorded data to probabilities \cite{White2,Haffner}. In this case, the standard QT requires a number of measurements which scales too fast with the dimension of the quantum system. For a system composed of $N$ qubits, for example, it is necessary to perform $6^N$ projections \cite{Altepeter}. In the case of using MUB-QT, the number of measurements adopted is minimal \cite{Wooters2,Ivanovic,Klimov08}. For this system of $N$ qubits, it is necessary to perform only
$2^N(2^N+1)$ projections.

The advent of new technological fields, such as quantum communication, is bringing more motivations for the development of QT techniques. The use of qudit states has also been shown to be relevant for these new applications \cite{Durt,Zeilinger} and, therefore, the development of techniques that allow better qudit state reconstructions is of upmost importance.

In this work we present a technique that allows the MUB-QT of photonic qudit states. The qudits are encoded using the linear transverse momentum of single photons transmitted through a diffractive aperture \cite{Leo2,Howell}. Since the qudit dimension is defined in terms of the number of available  paths for photon transmission at this aperture, we refer to them as spatial qudits. Our technique relies on the fact that spatial qudit states can be coherently modified by considering the capability of amplitude- and phase-modulation of programmable spatial light modulators (SLM) \cite{Glima2}. This capability, combined with a spatial filtering, allows the projection of the initial state onto the MUBs vectors. As a figure of merit for the performance of our reconstruction, we use the fidelity ($F$) \cite{Jozsa} of the obtained quantum states, which is defined as the overlap between the expected quantum state ($|\psi \rangle $) and the reconstructed one ($\rho$), $F\equiv\,\langle \psi |\rho |\psi \rangle$. For the 7- and 8-dimensional reconstructed qudit states, the fidelities obtained were greater than $90\%$.

\section{Setup description}

For a $D$-dimensional Hilbert space, whose dimension is a prime or a prime power number, there exist $D+1$ MUBs.  The constructive procedures for obtaining them have been explicitly given in \cite{Wooters2,Ivanovic,Klimov08}. The density operator of a qudit-$D$ is represented in terms of the MUBs by \cite{Ivanovic}
\begin{equation}\label{OpeMUB}
\rho=\sum_{\alpha=1}^{D+1}\sum_{m=1}^D p_m^{(\alpha)}\Pi_m^{(\alpha)}-I,
\end{equation} where $\Pi_m^{(\alpha)}\equiv  \ket{\psi_m^{(\alpha)}} \bra{\psi_m^{(\alpha)}}$, and
$p_m^{(\alpha)}=Tr(\rho\Pi_m^{(\alpha)})$ is the probability for
projecting $\rho$ onto the $m$-th state $|\psi_m^{(\alpha)}\rangle$
of the $\alpha$ MUB, with $\alpha=1,2,...,D+1$. The MUB-QT is performed by determining the $p_m^{(\alpha)}$ values from the experimental data acquired.

\begin{figure}[h]
\centerline{\includegraphics[width=0.7\textwidth]{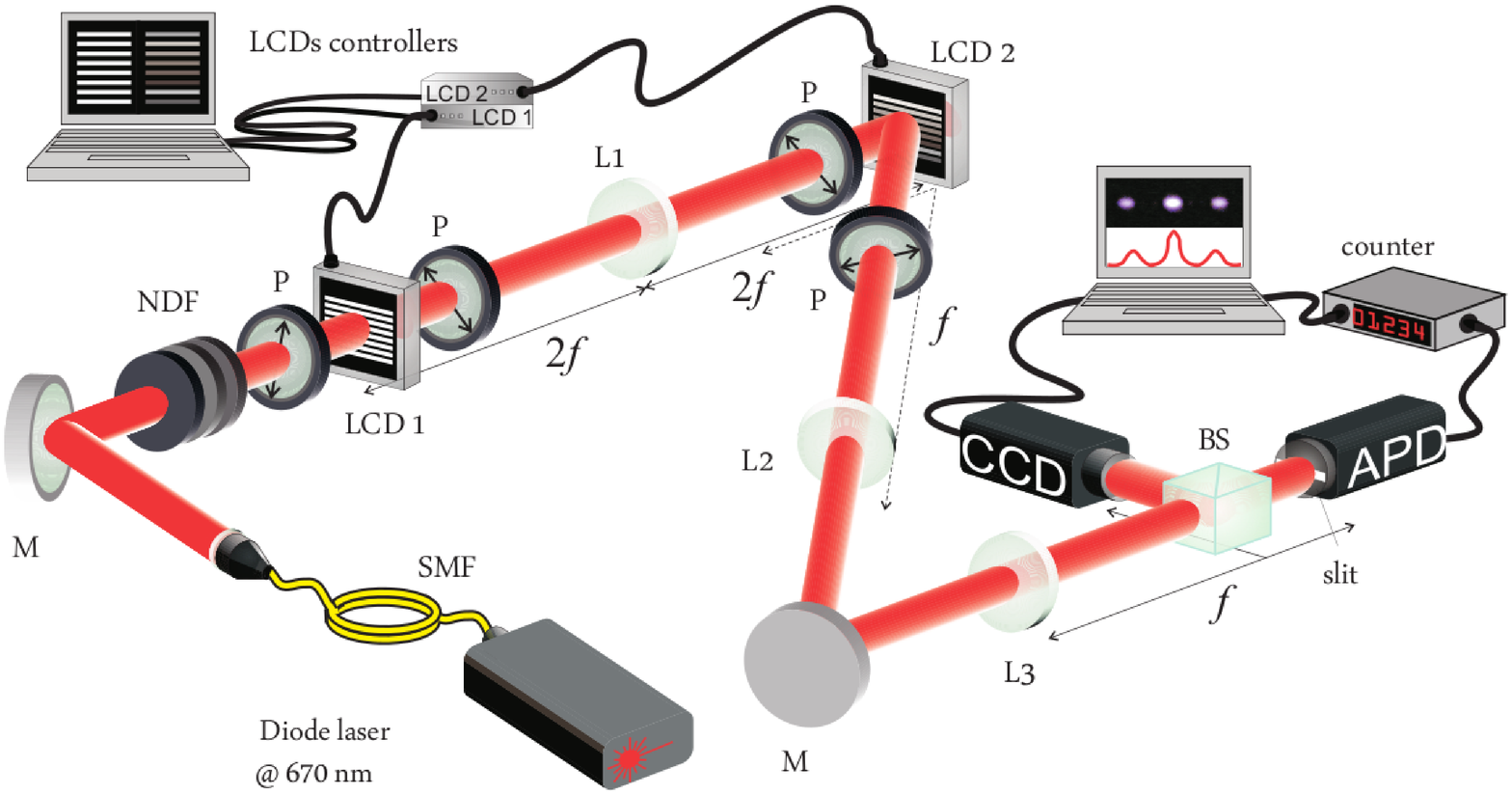}}
\vspace{-1.5cm}
\caption{Experimental Setup. See the main text for
details.} \label{Fig:Setup}
\end{figure}

The experimental setup used for the generation and reconstruction of the spatial qudit states is shown, schematically, in Fig.~\ref{Fig:Setup}. A pigtailed single mode diode laser, operating at $670$~nm, is extremely attenuated to the single photon level by using absorptive neutral density filters (NDF). After this, the photons are sent to a transmissive spatial light modulator, which is composed of two polarizers (P) and a twisted nematic liquid crystal display (LCD1) working for amplitude-only modulation. This spatial light modulator is addressed with a $D$-multi slit as shown in Fig.~\ref{Fig:Setup}. The slit's width is $2a=104$~$\mu$m and the distance between two consecutive slits is $d=208$~$\mu$m. The state of the single photon transmitted through the SLM can be written as \cite{Leo2,Glima2}
\begin{equation} \label{State}
\ket{\Psi} = \sum_{l=-l_D}^{l_D} \beta_l \ket{l},
\end{equation} where $l_D=(D-1)/2$ and the state $\ket{l}$ is a single photon state defined, up to a global phase
factor, as
\begin{equation}
\ket{l} \equiv \sqrt{\frac{a}{\pi}} \itgf{q} e^{-iqld}\sinc(qa)\ket{1q}.
\end{equation} It represents the state of the photon transmitted by the $l$th-slit of the SLM. $\ket{1q}$ is the Fock state of a photon with the transverse wave vector $\vec{q}$. The $\ket{l}$ states form the logical basis in the $D$-dimensional Hilbert space of the transmitted photons, whose dimension $D$ is defined by the number of slits addressed in the SLM. In this work we considered the generation of $7$- and $8$-dimensional spatial qudit states. Higher dimensional qudits can, in principle, also be generated by considering more complexes apertures. The $\beta_l$ coefficients are dependent on the spatial profile and the wavefront curvature of the laser beam at the SLM plane. The first SLM can also be used to control each slit transmission independently, such that the initial state of Eq.~(\ref{State}) can be modified to
\begin{equation}
\ket{\Psi}_{mod1} = \frac{1}{\sqrt{N}}
\sum_{l=-l_D}^{l_D} \lambda_l \beta_l \ket{l},
\end {equation}
where \protect{$\lambda_l\equiv\sqrt{t_l}$}, with $t_l$ representing the transmission of the $l$-slit, and $N$ is the normalization constant given by $N \equiv \sum{\lambda_l^2 \beta_l^2}$ \cite{Glima2}.

After being transmitted by the SLM, the modulated beam is imaged with a $7.5$~cm focal length lens (L1) onto a reflective SLM, configured for doing phase-only modulation. This is done in a $4f$-configuration, such that the LCD of the second modulator is at a distance of $30$~cm from the first LCD. This SLM is used to modulate the phase of the beam, independently, at each point of slit image formation. At the plane of image formation, the diffracted photons can again be described as spatial qudit states \cite{GLima3}. Thus, the phase modulation being done by the second SLM allows the control of the imaginary parts of the qudit states. The state $\ket{\Psi}_{mod1}$ can then be modified to
\begin{equation}
\ket{\Psi}_{mod2} = \frac{1}{\sqrt{N}} \sum_{l=-l_D}^{l_D} \lambda_l  \beta_l e^{i\theta_l}\ket{-l},
\end{equation} where $\theta_l$ is the phase given at the image of the $l$-th slit. Therefore, the two SLMs allow for a full manipulation of the initial spatial qudit states.

After the reflection at the second SLM, the beam is collimated by a $25$~cm focal length lens L$_2$, such that L$_1$ and L$_2$ form a telescope with a magnification factor of $3.3$. The collimated beam is then focused by a $1$~m focal length lens ($L_3$), and a point-like detector (APD) is used to record the single photons at the center of the interference pattern formed in the Fourier transform plane. The CCD camera, shown in Fig.~\ref{Fig:Setup}, was used just for the initial alignment of the SLMs without attenuating the laser beam. The point-like detector is composed of a conventional avalanche photo-counting module with a slit of $20$~$\mu$m in front of it. The spatial filtering being introduced by the point-like detector is the last ingredient necessary for projecting the spatial qudit states onto the MUBs' vectors \cite{Glima,Japas}. The single count rate of detecting the photon at transversal position $x$ at the focal plane of lens $L_3$ ($z_m-$plane) is given by:
\begin{eqnarray}
C (x)& \propto & \left\vert \left\langle \text{vac}\right\vert E_{s}^{(+)}(x,z_{m})\left\vert \psi \right\rangle_{mod2} \right\vert ^{2} \notag \\
                             & \propto & \text{sinc}^{2}\left( \frac{kxa}{f_3}\right) \left\vert \sum_{l}\beta_{l}\lambda _{l}e^{i\theta_{l} }e^{i\frac{ldkx}{f_3}} \right\vert^{2}
\end{eqnarray} where we have used that $E^{(+)}(x,z_m) \propto \hat{a}_{kx/f_3}$ and $ \left\langle \text{vac}  \right\vert \hat{a}_{kx/f_{3}} \ket{l} = \text{sinc}\left(kxa/f_3\right) e^{-i\frac{ldkx}{f_3}}$. $f_3$ is the focal length of lens $L_3$. In our case, when the point-like detector is fixed at the center of the interference pattern formed, the single count rate will be
\begin{equation}
C (0) \propto  \left\vert \sum_{l}\beta_{l}\lambda _{l}e^{i\theta_{l}} \right\vert^{2} .
\end{equation}

By choosing the amplitude and phase modulation parameters as $\lambda _{l}=\epsilon_{m,l}^{(\alpha)}$ and $\theta_{l}=\varphi_{m,l}^{(\alpha)}$, respectively, we get that the single count rate at the center of the interference pattern is given by $C_m^{(\alpha)} \propto \left\vert \langle \psi_m^{(\alpha)} \vert \Psi \rangle \right \vert^{2}$, where the $m$-th spatial vector of the $\alpha$-th MUB is given by
\begin{equation} \label{Eq:transf1}
|\psi_m^{(\alpha)}\rangle=\sum_l\epsilon_{m,l}^{(\alpha)}e^{-i\varphi_{m,l}^{(\alpha)}}|l\rangle.
\end{equation}

Since in the MUB-QT an overcomplete set of measurements is considered, $C^{(\alpha)}_{m}$ can be used to determine the probability $p^{(\alpha)}_{m}$ of projecting the initial state $|\Psi\rangle$ onto the $m$-th projector of the $\alpha$-th MUB. This is done by changing the gray level of the apertures modulated in the first and the second LCD(s), respectively. For an arbitrary initial quantum state $\rho$, using the linear properties of quantum theory, we have that $C^{(\alpha)}_{m} \propto \text{Tr} (\rho \Pi_m^{(\alpha)})$.

It is worth to mention that when $D$ is a prime number, all the probabilities for a given MUB, can also be obtained from a
single interference pattern formed by a specific phase modulation of the second SLM. In this case, all the probabilities
of the MUB considered can be inferred by measuring the single counts along the transversal direction $x$. This means
that with a single experimental configuration one can get, \emph{only in this specific case}, all the projective measurements of
a given standard MUB labeled by the index $\alpha$.

\begin{figure}[t]
\centerline{\includegraphics[width=0.65\textwidth]{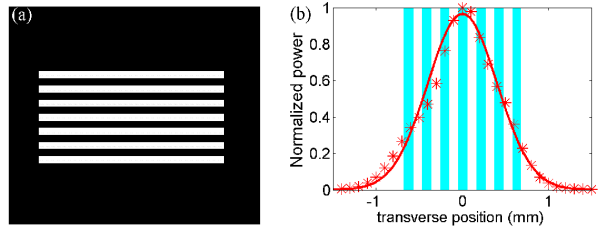}}
\vspace{-2.5cm}
\caption{The initial qudit-$7$ state. In (a) one can see the aperture modulated at the first SLM. In (b) there is a comparison between the laser beam spatial profile and this aperture.} \label{Fig:7quditIState}
\end{figure}

\section{Experimental results}

\subsection{qudit-$7$ state preparation}

We now present the experimental results obtained for the MUB-QT of a spatial qudit-$7$ state. The qudit-$7$ state was generated with the laser beam well collimated after the optical fiber, such that the first SLM can be considered to be approximately at the waist of the beam. In this case the initial qudit state is supposed to have only real coefficients. The values of $\beta_l$ are determined by the spatial distribution of the beam and by the aperture addressed in the SLM1.
The apperture modulated in this SLM is shown in Fig.~\ref{Fig:7quditIState}(a). In Fig.~\ref{Fig:7quditIState}(b) there is a comparison between the laser beam spatial profile and this aperture.
For such configuration, we can expect that the spatial qudit state being generated in the experiment is described by
$\ket{\Psi_7}_{expc}=0.256\ket{-3}+0.362\ket{-2}+0.443\ket{-1}+0.473\ket{0}+0.439\ket{1}+0.352\ket{2}+0.254\ket{3}$.

\begin{figure}[b]
\vspace{0.5cm}
\centerline{\includegraphics[width=0.7\textwidth]{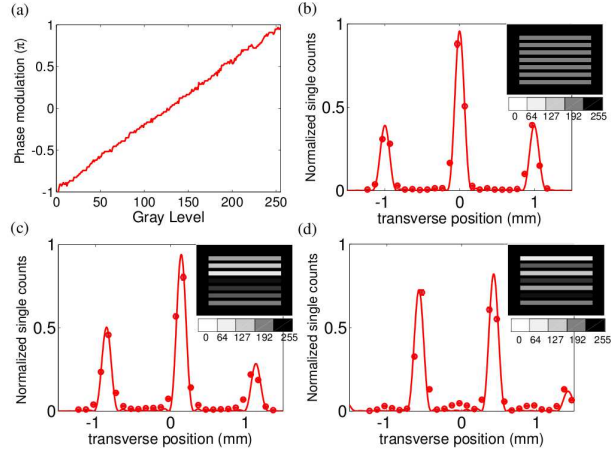}}
\vspace{-0.2cm}
\caption{Phases modulated at the second SLM and the corresponding recorded interference patterns. Figure (a)  illustrates the relation between the gray level of the second LCD and the phase being modulated by this modulator. Some of the modulations used in the MUB-QT, for projecting the generated qudit-$7$ state onto $\ket{\psi^{\alpha}_m}$, are shown in the insets of figures (b), (c) and (d). In (b) $\alpha=7$ and $m=1$. In (c) $\alpha=7$ and $m=2$, and in (d) $\alpha=7$ and $m=5$. The corresponding recorded interference patterns (points) are compared with the expected ones (lines). The expected patterns are calculated from the amplitudes of $\ket{\Psi_7}_{expc}$. The integration time of these measurements was one second and the maximal single count rate recorded was 10000 counts/s.} \label{Fig:7quditPM}
\end{figure}

\subsection{qudit-$7$ state reconstruction}

\begin{figure}[t]
\centerline{\includegraphics[width=0.7\textwidth]{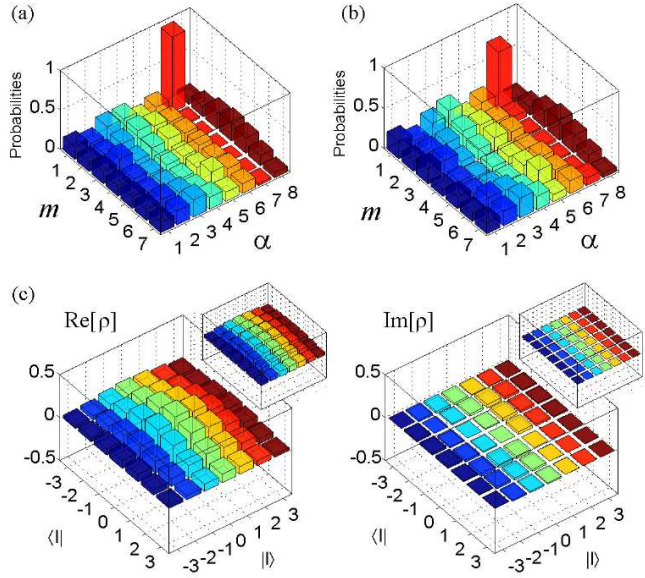}}
\vspace{-5.0cm}
\caption{MUB-QT of the generated qudit-$7$ state. (a) The expected probabilities based on the predicted state $\ket{\Psi_7}_{expc}$. (b) The recorded probabilities with single counts. (c) The real and the imaginary parts of the reconstructed state. On the insets of (c) the parts of the expected density operator are shown.} \label{Fig:7quditRecons}
\end{figure}

The MUBs vectors for a seven dimensional Hilbert space are given explicitly by Eq. (11) in \cite{Wooters2}, and in our case they are obtained by considering at Eq.~(\ref{Eq:transf1}) that
\begin{equation}
\epsilon_{m,l}^{(\alpha)}=\frac{1}{\sqrt{7}}, \,\,\,
\varphi^{(\alpha)}_{m,l}   = \frac{2 \pi (\alpha l^2 + m l)}{7},
\end{equation} with $\alpha=1,2,...,7$ and $m,l=-3,...,3$. $\varphi^{(\alpha)}_{m,l}$ is computed with $\textrm{mod} (2 \pi)$. The probabilities for projecting onto the logical base vectors, which form one of the MUBs for a $7$-dimensional Hilbert space, can be determined by measuring the single counts at the plane of image formation. This is done by removing the second SLM and scanning transversely the APD at this plane \cite{GLima3}. To generate and measure the projection probabilities for all the other seven dimensional MUBs vectors, it is sufficient to consider only the phase modulation implemented by the second SLM and the detection at the Fourier transform plane as mentioned above. In Fig.~\ref{Fig:7quditPM}(a), one can see the dependency of the phase being modulated and the gray level of the LCD2. At Fig.~\ref{Fig:7quditPM} it is also shown the modulations used to project the state $\ket{\Psi_7}_{expc}$ onto some of the vectors of the $7$-th MUB. The corresponding interference patterns were completely recorded just to show the purity of the modified states. The visibility of these patterns guarantees that the modulation is not introducing decoherence to the qudit states, showing that the modified states are nearly pure \cite{Glima2}. In Fig.~\ref{Fig:7quditRecons}(a) and Fig.~\ref{Fig:7quditRecons}(b) there is a comparison between the recorded spatial projection probabilities and the expected ones. The expected probabilities are calculated from the above given expression for $\ket{\Psi_7}_{expc}$. One can clearly see that the obtained probabilities are very close to the expected ones.

\begin{figure*}[t]
\centerline{\includegraphics[width=1.0\textwidth]{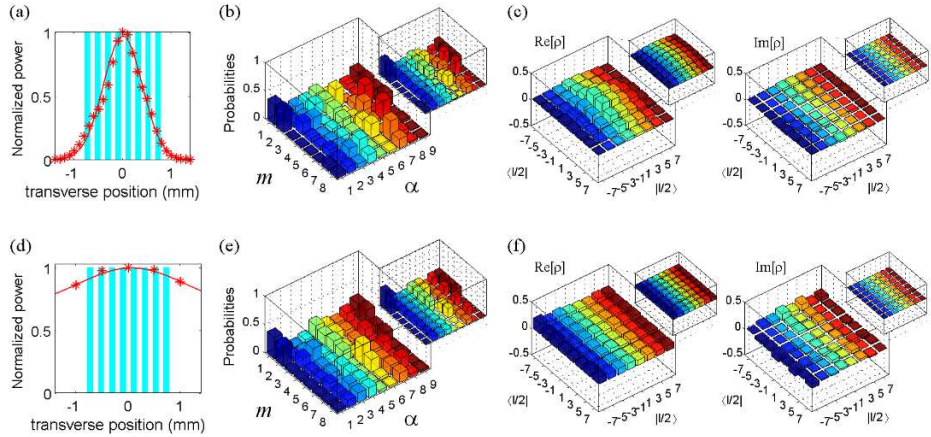}}
\vspace{-3.0cm} \caption{Generation and reconstruction of the spatial qudit-$8$ states. In (a) [(d)] there is a comparison between the object being modulated at the first SLM and the spatial
laser profile used for generating the first (second) qudit-$8$ state. In (b) and (e) one can see a comparison between the recorded probabilities and the expected ones (insets) that are calculated from the states $\ket{\Psi^{(1)}_{8}}_{expc}$ and $\ket{\Psi^{(2)}_{8}}_{expc}$, respectively. In (c) [(f)] the corresponding reconstructed density operator for the first (second) qudit-$8$ state generated is shown. On the insets of (c) and (f) one can see the expected density operators.} \label{Fig:8quditRecons}
\end{figure*}

From the measured probabilities one can reconstruct the density operator using Eq.~(\ref{OpeMUB}). However, due to the inherent statistical fluctuations of the experimental results, the obtained density operator may not be positive semidefinite \cite{James}. Thus, it is necessary to optimize it by using numerical techniques. For the states measured in our experiment we followed the ``Forced-purity'' approach, which was shown to provide sufficient optimization for the density operators of low-entropy high-dimensional qudits \cite{James2}. This is exactly our case and the qudit-$7$ obtained density operator is graphically compared with the expected one in Fig.~\ref{Fig:7quditRecons}(c). The expected real and imaginary parts are shown on the insets. The fidelity \cite{Jozsa} between these states was $F_{7q}=0.96\pm0.03$. This error was calculated taking into account the Poissonian distribution for the single counts recorded in the experiment. The errors on the elements of the reconstructed state can be calculated directly from Eq.~(\ref{OpeMUB}). Experimental obtained fidelities are usually limited by the imperfections of an experimental setup. We believe that in our case, these imperfections are mainly due to the mismatch that can occur while superposing the image of the first SLM onto the image modulated on the LCD of the second SLM.

\subsection{Preparing the qudit-$8$ states}

In our work we also considered the generation and
reconstruction of two distinct spatial qudit-$8$ states. These states were generated by changing the spatial distribution of the collimated laser beam at the plane of the first SLM. This was done with an optical beam expander mounted before the ND filters. The two spatial profiles generated for each state are shown in Fig.~\ref{Fig:8quditRecons}(a) and Fig.~\ref{Fig:8quditRecons}(d), respectively, where they are also compared with the object (the $8$-multi slit) modulated at the first SLM. From these comparison we expected the first state generated to be: $\ket{\Psi^{(1)}_{8}}_{expc}=0.217\ket{-\frac{7}{2}}+0.308\ket{-\frac{5}{2}}+0.399\ket{-\frac{3}{2}}+0.456\ket{-\frac{1}{2}}+0.453\ket{\frac{1}{2}}+0.393\ket{\frac{3}{2}}+0.297\ket{\frac{5}{2}}+0.202\ket{\frac{7}{2}}$, and the second one: $\ket{\Psi^{(2)}_{8}}_{expc}=0.343\ket{-\frac{7}{2}}+0.350\ket{-\frac{5}{2}}+0.355\ket{-\frac{3}{2}}+0.358\ket{-\frac{1}{2}}+0.359\ket{\frac{1}{2}}+0.357\ket{\frac{3}{2}}+0.354\ket{\frac{5}{2}}+0.348\ket{\frac{7}{2}}$.

The case of a spatial qudit-$8$ state has many interesting features.
The first one is that it can be thought as a composite system of
three spatial qubits.  This can easily be seen by considering the
labeling of the $\ket{l}$ states as follows:
$\ket{-\frac{7}{2}}\rightarrow\ket{000}$,
$\ket{-\frac{5}{2}}\rightarrow\ket{001}$,
$\ket{-\frac{3}{2}}\rightarrow\ket{010}$,
$\ket{-\frac{1}{2}}\rightarrow\ket{011}$,
$\ket{\frac{1}{2}}\rightarrow\ket{100}$,
$\ket{\frac{3}{2}}\rightarrow\ket{101}$,
$\ket{\frac{5}{2}}\rightarrow\ket{110}$,
$\ket{\frac{7}{2}}\rightarrow\ket{111}$. These spatial qubits can, therefore, be used to simulate and study entangled states of three qubits. This is the case of the state $\ket{\Psi^{(1)}_{8}}_{expc}$ which can not be factorized. The state
$\ket{\Psi^{(2)}_{8}}_{expc}$ is almost a product state. In general, in case of having a $2^N$ dimensional qudit, with $N$ integer, this experimental setup allows us for mimicking a multipartite system of $N$ qubits. These subsystems are encoded on the transverse modes of the transmitted photon
and, therefore, they cannot be used to test the quantum non-locality \cite{EPR}. Nevertheless, they can be used for studying new techniques proposed for characterizing the quantum correlations of high-dimensional quantum systems \cite{Reinaldo,Glima10}, or also for studying quantum information processes \cite{Spreeuw,Steve2,Graci}.

\subsection{Reconstruction of the qudit-$8$ states}

Another interesting aspect of considering an qudit-$8$ is the fact that there are four distinct classes of MUBs for an eight-dimensional Hilbert space \cite{Klimov08,Romero05}. The spatial qudit-$8$ can, therefore, be reconstructed by considering another type of projective measurements, that correspond to the projections onto the MUBs vectors of a new family of MUBs. To shown the flexibility of our setup, we considered the MUBs given in Table I of \cite{Klimov08}. All the bases in this table requires one conditional gate acting on mimicked qubits. The corresponding MUB bases are obtained by applying a unitary transformation on the standard computational basis. These bases have the interesting properties of being optimal, when dealing with physical subsystems, since they require a lower number of controlled-NOT operations to be built. For sake of completeness, we have included all these unitary transformations in Appendix \ref{App:UT}. From these transformations we get the values of the $\{\epsilon^{(\alpha)}_{m,l},\varphi^{(\alpha)}_{m,l}\}$ parameters by using that $\epsilon^{(\alpha)}_{m,l}=\vert U^{(\alpha)}_{m,l} \vert$ and $\varphi^{(\alpha)}_{m,l}=\textrm{arg}(U^{(\alpha)}_{m,l})$, with $\alpha=1,2,...,8$ and $m,l=-7/2,-5/2,...,7/2$. For generating these spatial projections it was necessary to change the transmission of the slits addressed at the first SLM.

The probabilities recorded are shown in Fig.~\ref{Fig:8quditRecons}(b) and Fig.~\ref{Fig:8quditRecons}(e), where they are also compared to the expected ones (insets) calculated from the states $\ket{\Psi^{(1)}_{8}}_{expc}$ and $\ket{\Psi^{(2)}_{8}}_{expc}$, respectively. The corresponding reconstructed density operator for the first (second) qudit-$8$ state generated is shown in Fig.~\ref{Fig:8quditRecons}(c) [Fig.~\ref{Fig:8quditRecons}(f)]. The fidelities of the reconstructed states with the expected ones [Shown in the insets of Fig.~\ref{Fig:8quditRecons}(c) and Fig.~\ref{Fig:8quditRecons}(f)], after numerical optimization, were $F^{(1)}_{8q}=0.93\pm0.03$ and $F^{(2)}_{8q}=0.91\pm0.03$, respectively.

\section{Conclusion}

In conclusion, we have created an experimental setup based on the use of SLMs which allows the generation and the reconstruction of distinct types of high-dimensional photonic states based on the mutually unbiased basis. Experimental results for the reconstruction of seven and eight dimensional systems were presented. The possibility of extending our study to even higher dimensional systems sounds feasible. For this purpose, the multi-slit arrays modulated onto the SLM(s), need to be replaced by more complexes apertures which will naturally define higher dimensional states.

Furthermore, the presented experimental setup can be slightly modified for investigating other types of quantum state reconstruction of higher dimensional quantum systems. For instance, it can be used for the quantum tomography based on the informationally complete bases of Ref \cite{Caves04}, unbiassed non-orthogonal bases \cite{Sainz10}, or also for the QT that is based on the equidistant states \cite{Paiva10}.

Therefore, this work can be seen as fundamental investigation which may allow more versatility for the experimental studies of high-dimensional quantum systems.

\appendix
\section{Unitary transformations used for the qudit-8 MUB-QT}
\label{App:UT}

We give the expressions for the unitary operations acting onto the logical basis $\ket{l}$ of Eq.~(\ref{State}), which define the new family of MUBs used for the qudit-8 MUB-QT \cite{Klimov08}.

\begin{equation*}
U^{\left( 1\right) } =\frac{1}{\sqrt{8}}\left(
\begin{array}{rrrrrrrr}
1 & -i & -1 & i & -1 & i & 1 & -i \\
-i & 1 & i & -1 & i & -1 & -i & 1 \\
-i & 1 & -i & 1 & i & -1 & i & -1 \\
1 & -i & 1 & -i & -1 & i & -1 & i \\
1 & -i & -1 & i & 1 & -i & -1 & i \\
-i & 1 & i & -1 & -i & 1 & i & -1 \\
-i & 1 & -i & 1 & -i & 1 & -i & 1 \\
1 & -i & 1 & -i & 1 & -i & 1 & -i%
\end{array} \right).
\end{equation*}

\begin{equation*}
U^{\left( 2\right) } =\frac{1}{\sqrt{8}}\left(
\begin{array}{rrrrrrrr}
1 & -1 & -1 & 1 & -i & i & i & -i \\
1 & 1 & -1 & -1 & -i & -i & i & i \\
-1 & 1 & -1 & 1 & i & -i & i & -i \\
1 & 1 & 1 & 1 & -i & -i & -i & -i \\
-i & i & i & -i & 1 & -1 & -1 & 1 \\
-i & -i & i & i & 1 & 1 & -1 & -1 \\
i & -i & i & -i & -1 & 1 & -1 & 1 \\
-i & -i & -i & -i & 1 & 1 & 1 & 1%
\end{array} \right).
\end{equation*}

\begin{equation*}
U^{\left( 3\right) } =\frac{1}{\sqrt{2}}\left(
\begin{array}{rrrrrrrr}
1 & 0 & -1 & 0 & 0 & 0 & 0 & 0 \\
0 & i & 0 & -i & 0 & 0 & 0 & 0 \\
0 & 1 & 0 & 1 & 0 & 0 & 0 & 0 \\
i & 0 & i & 0 & 0 & 0 & 0 & 0 \\
0 & 0 & 0 & 0 & 1 & 0 & -1 & 0 \\
0 & 0 & 0 & 0 & 0 & i & 0 & -i \\
0 & 0 & 0 & 0 & 0 & 1 & 0 & 1 \\
0 & 0 & 0 & 0 & i & 0 & i & 0%
\end{array} \right).
\end{equation*}

\begin{equation*}
U^{\left( 4\right) } =\frac{1}{2}\left(
\begin{array}{rrrrrrrr}
1 & -1 & 0 & 0 & -1 & -1 & 0 & 0 \\
-1 & 1 & 0 & 0 & -1 & -1 & 0 & 0 \\
0 & 0 & 1 & -1 & 0 & 0 & -1 & -1 \\
0 & 0 & -1 & 1 & 0 & 0 & -1 & -1 \\
1 & 1 & 0 & 0 & -1 & 1 & 0 & 0 \\
1 & 1 & 0 & 0 & 1 & -1 & 0 & 0 \\
0 & 0 & 1 & 1 & 0 & 0 & -1 & 1 \\
0 & 0 & 1 & 1 & 0 & 0 & 1 & -1%
\end{array} \right).
\end{equation*}

\begin{equation*}
U^{\left( 5\right) } =\frac{1}{\sqrt{8}}\left(
\begin{array}{rrrrrrrr}
1 & -i & -i & -1 & -1 & -i & i & -1 \\
-i & 1 & -1 & -i & -i & -1 & -1 & i \\
-i & -1 & 1 & -i & i & -1 & -1 & -i \\
-1 & -i & -i & 1 & -1 & i & -i & -1 \\
-i & 1 & -1 & -i & i & 1 & 1 & -i \\
1 & -i & -i & -1 & 1 & i & -i & 1 \\
-1 & -i & -i & 1 & 1 & -i & i & 1 \\
-i & -1 & 1 & -i & -i & 1 & 1 & i
\end{array} \right).
\end{equation*}

\begin{equation*}
U^{\left( 6\right) } =\frac{1}{2} \left(
\begin{array}{rrrrrrrr}
1 & 0 & -1 & 0 & -1 & 0 & 1 & 0 \\
0 & i & 0 & -i & 0 & -i & 0 & i \\
1 & 0 & 1 & 0 & -1 & 0 & -1 & 0 \\
0 & i & 0 & i & 0 & -i & 0 & -i \\
0 & 1 & 0 & -1 & 0 & 1 & 0 & -1 \\
i & 0 & -i & 0 & i & 0 & -i & 0 \\
0 & 1 & 0 & 1 & 0 & 1 & 0 & 1 \\
i & 0 & i & 0 & i & 0 & i & 0
\end{array} \right).
\end{equation*}

\begin{equation*}
U^{\left( 7\right) } =\frac{1}{2}\left(
\begin{array}{rrrrrrrr}
1 & -i & 0 & 0 & -1 & i & 0 & 0 \\
-i & 1 & 0 & 0 & i & -1 & 0 & 0 \\
0 & 0 & 1 & -i & 0 & 0 & -1 & i \\
0 & 0 & -i & 1 & 0 & 0 & i & -1 \\
0 & 0 & i & 1 & 0 & 0 & i & 1 \\
0 & 0 & 1 & i & 0 & 0 & 1 & i \\
i & 1 & 0 & 0 & i & 1 & 0 & 0 \\
1 & i & 0 & 0 & 1 & i & 0 & 0%
\end{array} \right).
\end{equation*}

\begin{equation*}
U^{\left( 8\right) } =\frac{1}{\sqrt{8}}\left(
\begin{array}{rrrrrrrr}
1 & -1 & -1 & 1 & -1 & 1 & -1 & 1 \\
1 & 1 & -1 & -1 & -1 & -1 & -1 & -1 \\
-1 & 1 & 1 & -1 & -1 & 1 & -1 & 1 \\
-1 & -1 & 1 & 1 & -1 & -1 & -1 & -1 \\
1 & -1 & 1 & -1 & -1 & 1 & 1 & -1 \\
1 & 1 & 1 & 1 & -1 & -1 & 1 & 1 \\
1 & -1 & 1 & -1 & 1 & -1 & -1 & 1 \\
1 & 1 & 1 & 1 & 1 & 1 & -1 & -1%
\end{array} \right).
\end{equation*}

\begin{equation*}
U^{\left( 9\right) } =\frac{1}{2}\left(
\begin{array}{rrrrrrrr}
1 & 0 & -i & 0 & -1 & 0 & i & 0 \\
0 & 1 & 0 & -i & 0 & -1 & 0 & i \\
-i & 0 & 1 & 0 & i & 0 & -1 & 0 \\
0 & -i & 0 & 1 & 0 & i & 0 & -1 \\
-i & 0 & 1 & 0 & -i & 0 & 1 & 0 \\
0 & -i & 0 & 1 & 0 & -i & 0 & 1 \\
1 & 0 & -i & 0 & 1 & 0 & -i & 0 \\
0 & 1 & 0 & -i & 0 & 1 & 0 & -i
\end{array} \right).
\end{equation*}

Here, the index $\alpha$ in $U^{(\alpha)}$ denotes a specific mutually unbiased basis in the $8$-dimensional Hilbert space. It can be noted that in transformations with $\alpha=1,2,5,8$, all the $\epsilon^{(\alpha)}_{m,l}$ are equals to $1/\sqrt{8}$. Besides, in transformations with $\alpha=4,6,7,8$ we get that $\epsilon^{(\alpha)}_{m,l}=0$ and $1/2$ and finally for $\alpha=3$ that $\epsilon^{(\alpha)}_{m,l}=0$ and $1/\sqrt{2}$. These amplitude coefficients are implemented with the SLM$_1$. Furthermore, the values for the phase modulation at the SLM$_2$ [Eq. (\ref{Eq:transf1})], are given by $\varphi^{(\alpha)}_{m,l}=\textrm{arg}(U^{(\alpha)}_{m,l})$. These values belong to the discrete set of $\{0,\pi/2,\pi,3\pi/2\}$. Hence the LCD must allow for full modulation at the $\{0,2\pi\}$ domain, which is effectively provided by the used SLM as it is shown in Fig. \ref{Fig:7quditPM}(a).

\section*{Acknowledgments}

This work was supported by Grants CONICYT~PFB08-024, PBCT PDA-25, FONDECYT~11085055 and FONDECYT~11085057. A. V. acknowledges grant DI09-0045 of Universidad de La Frontera.


\begin{thebibliography}{99}

\bibitem{Fano} U. Fano, ``Description of States in Quantum Mechanics by Density Matrix and Operator Techniques,'' Rev. Mod. Phys. \textbf{29}, 74–-93 (1957).

\bibitem{James} D. F. V. James, P. G. Kwiat, W. J. Munro, and A. G. White, ``Measurement of qubits,'' \pra \textbf{64}, 052312 (2001).

\bibitem{Thew} R. T. Thew, K. Nemoto, A. G. White, and W. J. Munro, ``Qudit quantum-state tomography,'' \pra \textbf{66}, 012303 (2002).

\bibitem{Langford} M. D de Burgh,N.K. Langford, A. C. Doherty, and A. Gilchrist, ``Choice of measurement sets in qubit tomography,'' \pra \textbf{78}, 052122 (2008).

\bibitem{Wooters2}  W. K. Wootters and B. D. Fields, ``Optimal state-determination by mutually unbiased measurements,'' Ann. Phys \textbf{191}, 363--381 (1989).

\bibitem{Steinberg} R. B. A. Adamson and A. M. Steinberg, ``Improving Quantum State Estimation with Mutually Unbiased Bases,'' \prl \textbf{105}, 030406 (2010).

\bibitem{Haffner} H. Haffner, W. Hansel, C. F. Roos, J. Benhelm, D. Chek-al-kar, M. Chwalla, T. Korber, U. D. Rapol, M. Riebe, P. O. Schmidt, C. Becher, O. Guhne, W. Dur, R. Blatt, ``Scalable multiparticle entanglement of trapped ions,'' Nature \textbf{438}, 643--646 (2005).

\bibitem{White2}  N. K. Langford, R. B. Dalton, M. D. Harvey, J. L. O'Brien, G. J. Pryde, A. Gilchrist, S. D. Bartlett, and A. G. White, ``Measuring Entangled Qutrits and Their Use for Quantum Bit Commitment,'' \prl \textbf{93}, 053601 (2004).

\bibitem{Altepeter} J. B. Altepeter, E. R. Jeffrey, and P. G. Kwiat, ``Chap.3:Photonic State Tomography,'' Advances in AMO Physics, Vol. 52 (Elsevier, 2006).

\bibitem{Ivanovic} I. D. Ivanovic, ``Geometrical description of quantal state determination,'' J. Phys. A \textbf{14}, 3241--3245 (1981).

\bibitem{Klimov08} A. B. Klimov, C. Muñoz, A. Fern\'{a}ndez and C. Saavedra, ``Optimal quantum-state reconstruction for cold trapped ions,'' Phys. Rev. A \textbf{77}, 060303(R) (2008).

\bibitem{Durt} T. Durt, D. Kaszlikowski, J. L. Chen, and L. C. Kwek, ``Security of quantum key distributions with entangled qudits,'' Phys. Rev. A {\bf 69}, 032313 (2004).

\bibitem{Zeilinger} D. Kaszlikowski, P. Gnacin´ski, M. Zukowski, W. Miklaszewski and A. Zeilinger, ``Violations of Local Realism by Two Entangled N-Dimensional Systems Are Stronger than for Two Qubit,'' Phys. Rev. Lett. {\bf 85}, 4418--4421 (2000).

\bibitem{Leo2} L. Neves, G. Lima, J. G. A. G\'omez, C. H. Monken, C. Saavedra, and S. P\'adua, ``Generation of Entangled States of Qudits using Twin Photons,''  \prl {\bf 94}, 100501 (2005).

\bibitem{Howell} M. N. O.-Hale, I. A. Khan, R. W. Boyd and J. C. Howell, ``Pixel Entanglement: Experimental Realization of Optically Entangled $d=3$ and $d=6$ Qudits,'' \prl \textbf{94}, 220501 (2005).

\bibitem{Glima2} G. Lima, A. Vargas, L. Neves, R. Guzm\'an, and C. Saavedra, ``Manipulating spatial qudit states with programmable optical devices,'' Opt. Express {\bf 17}, 10688--10696 (2009).

\bibitem{Jozsa} R. Jozsa, ``Fidelity for Mixed Quantum States,'' J. Mod. Opt. \textbf{41}, 2315-–2323 (1994).

\bibitem{GLima3} G. Lima, L. Neves, I. F. Santos, J. G. Aguirre G\'omez, C. Saavedra, and S. P\'adua, ``Propagation of spatially entangled qudits through free space,'' \pra {\bf 73}, 032340 (2006).

\bibitem{Glima} G. Lima, F. A. Torres-Ruiz, L. Neves, A Delgado, C Saavedra and S. P\'{a}dua, ``Measurement of spatial qubits,'' J. Phys. B \textbf{41}, 185501 (2008).

\bibitem{Japas} G. Taguchi,  T. Dougakiuchi, N. Yoshimoto, K. Kasai, M. Iinuma, H.F. Hofmann, and Y. Kadoya, ``Measurement and control of spatial qubits generated by passing photons through double slits,'' \pra \textbf{78}, 012307 (2008).

\bibitem{James2}  M. S. Kaznady and D. F. V. James, ``Numerical strategies for quantum tomography: Alternatives to full optimization,'' \pra \textbf{79}, 022109 (2009).

\bibitem{EPR} A. Einstein, B. Podolsky and N. Rosen, ``Can Quantum-Mechanical Description of Physical Reality Be Considered Complete?,'' Phys. Rev. \textbf{47}, 777--780 (1935).

\bibitem{Reinaldo} T. O. Maciel and R. O. Vianna, ``Viable entanglement detection of unknown mixed states in low dimensions,'' \pra \textbf{80}, 032325 (2009).

\bibitem{Glima10} G. Lima, E. S. G\'omez, A. Vargas, R. O. Vianna, and C. Saavedra, ``Fast entanglement detection for unknown states of two spatial qutrits,'' \pra \textbf{82}, 012302 (2010).

\bibitem{Spreeuw}  R. J. C. Spreeuw, ``Classical wave-optics analogy of quantum-information processing,'' \pra {\bf 63}, 062302 (2001).

\bibitem{Steve2} S. P. Walborn, D. S. Lamelle, M. P. Almeida, and P. H. Souto Ribeiro, ``Quantum Key Distribution with Higher-Order Alphabets Using Spatially Encoded Qudits,'' \prl \textbf{96}, 090501 (2006).

\bibitem{Graci} G. Puentes, C. La Mela, S. Ledesma, C. Iemmi, J. P. Paz, and M. Saraceno, ``Optical simulation of quantum algorithms using programmable liquid-crystal displays,'' \pra \textbf{69}, 042319 (2004).

\bibitem{Romero05} J. L. Romero, G. B{o}rk, A.B. Klimov, and L.L. S\'anchez-Soto, ``Structure of the sets of mutually unbiased bases for N qubits,'' \pra \textbf{72}, 062310 (2005).

\bibitem{Caves04} J. M. Renes, R. Blume-Kohout, A. J. Scott, and C. M. Caves, ``Symmetric informationally complete quantum measurements,'' J. Math. Phys. \textbf{45}, 2171--2181 (2004).

\bibitem{Sainz10} I. Sainz, L. Roa, A. B. Klimov, ``Unbiased nonorthogonal bases for tomographic reconstruction,'' \pra \textbf{81}, 052114 (2010).

\bibitem{Paiva10} C. Paiva, E. Burgos-Inostroza, O. Jim\'enez, A. Delgado, ``Quantum tomography via equidistant states,'' \pra \textbf{82}, 032115 (2010).

\end{thebibliography}
\end{document}